\documentclass{aastex631}

\usepackage{soul}

\newcommand\HI{\textrm{H}\,\scriptstyle\mathrm{I}}

\begin{document}

\title{Search for neutrino emission from the Cygnus Bubble based on LHAASO $\gamma$-ray observations}

\author[0000-0002-4300-5130]{Wenlian Li$^{\dagger}$}
\affiliation{Tsung-Dao Lee Institute, Shanghai Jiao Tong University, 201210 Shanghai, China}

\author[0000-0001-8510-2513]{Tian-Qi Huang$^{\dagger}$}
\affiliation{Key Laboratory of Particle Astrophysics and Experimental Physics Division and Computing Center, Institute of High Energy Physics, Chinese Academy of Sciences, 100049 Beijing, China}
\affiliation{TIANFU Cosmic Ray Research Center, Chengdu, Sichuan, China}

\author[0000-0003-1639-8829]{Donglian Xu}
\affiliation{Tsung-Dao Lee Institute, Shanghai Jiao Tong University, 201210 Shanghai, China}
\affiliation{School of Physics and Astronomy, Shanghai Jiao Tong University, Key Laboratory for Particle Astrophysics and Cosmology (MoE), Shanghai Key Laboratory for Particle Physics and Cosmology, 200240 Shanghai, China}

\author[0000-0002-5963-4281]{Huihai He}
\affiliation{Key Laboratory of Particle Astrophysics and Experimental Physics Division and Computing Center, Institute of High Energy Physics, Chinese Academy of Sciences, 100049 Beijing, China}
\affiliation{TIANFU Cosmic Ray Research Center, Chengdu, Sichuan, China}
\affiliation{University of Chinese Academy of Sciences, 100049 Beijing, China}


\correspondingauthor{D.L. Xu}
\email{donglianxu@sjtu.edu.cn}

\def\thefootnote{$\dagger$}\footnotetext{These authors contributed equally to this work}\def\thefootnote{\arabic{footnote}}





\begin{abstract}

The Cygnus region, which contains massive molecular and atomic clouds and young stars, is a promising Galactic neutrino source candidate. Cosmic rays transport in the region can produce neutrinos and $\gamma$-rays. Recently, the Large High Altitude Air Shower Observatory (LHAASO) detected an ultrahigh-energy $\gamma$-ray bubble (Cygnus Bubble) in this region. Using publicly available track events detected by the IceCube Neutrino Observatory in 7 years of full detector operation, we conduct searches for correlated neutrino signals from the Cygnus Bubble with neutrino emission templates based on LHAASO $\gamma$-ray observations. No significant signals were found for any employed templates. With the 7 TeV $\gamma$-ray flux template, we set a flux upper limit of 90\% confidence level (C.L.) for the neutrino emission from the Cygnus Bubble to be $5.7\times10^{-13}\, \mathrm{TeV}^{-1}\mathrm{cm}^{-2}\mathrm{s}^{-1}$ at 5 TeV. 

\end{abstract}



\section{Introduction} 
\label{sec:intro}



Cosmic rays are high-energy astrophysical particles, primarily protons and atomic nuclei, while their origins have been a mystery for a century. Under the confinement of the Galactic magnetic field, the observed cosmic rays with energies up to several PeV are believed to originate from Galactic sources, called PeVatrons. 
Cosmic rays interact with the interstellar medium or the radiation field, generating both neutrinos (e.g., $\pi^{+}\rightarrow\mu^{+}+\nu_{\mu}$) and $\gamma$-rays (e.g., $\pi^{0}\rightarrow2\gamma$).
High energy electrons can also produce $\gamma$-rays through inverse Compton scattering. However, the cross section suffers more stringent Klein–Nishina suppression for $\gamma$-rays with energies above 100 TeV. Therefore, the coincidence between neutrinos or $\gamma$-rays ($>100$ TeV) and gas clumps will provide critical evidence for the identification of hadronic PeVatrons.


The Cygnus region is an active star forming area in our Galaxy and hosts various astrophysical sources, including massive young star clusters (YMCs, e.g., Cygnus OB2), pulsar wind nebulae (PWNe, e.g., TeV J2032+4130), and supernova remnants (SNRs, e.g., $\gamma$-Cygni). Fermi-LAT detected an excess of $\gamma$-ray emission (1-100 GeV) from the direction of the Cygnus region after subtracting the interstellar background and all known sources \citep{2011Sci...334.1103A}. The hard $\gamma$-ray spectrum points to freshly accelerated cosmic rays, whether they are cosmic ray electrons or nuclei. This $\sim 2^{\circ}$ extended $\gamma$-ray source, known as Cygnus Cocoon, has been further observed at TeV energies by ARGO-YBJ \citep{2014ApJ...790..152B} and HAWC \citep{2021NatAs...5..465A}. 
In the latest observation of the Cygnus region, LHAASO reported the Cygnus Bubble at ultra-high energy  \citep{cygnusbubble2023}, extending to more than $6^{\circ}$ from the core, which is much larger than the Cygnus Cocoon.
The $\gamma$-ray brightness follows the distribution of the molecular gas, especially for $\gamma$-rays above 100 TeV, suggesting that these $\gamma$-rays are produced by the collision between the gas and the cosmic rays. We expect to observe high-energy neutrinos from the Cygnus Bubble.

The IceCube Neutrino Observatory has previously searched for neutrino emission from the PeVatron candidates observed by LHAASO. The hadronic components of the Crab Nebula and LHAASO J1849-0003 are constrained to be no more than $\sim 80\%$ and $\sim90\%$ of the total $\gamma$-rays observed \citep{2022MNRAS.514..852H, 2023ApJ...945L...8A}. 
\textcolor{black}{As for the Cygnus region, the hadronic contribution is contained to be less than 60\%  \citep{2019ICRC...36..932K}, while the resolved sources (e.g., TeV J2032+4130) are not removed from this region.}
\textcolor{black}{\cite{2022ApJ...930L..24A} observed neutrinos from the direction of the X-ray binary Cyg X-3 with a pretrial p-value of 0.009 ($2.4\sigma$) in the time-integrated search.}
\textcolor{black}{Recently, \cite{Neronov:2023hzu} claimed a $3\sigma$ excess of neutrino signals from the central region ($\sim 1^{\circ}$) of the Cygnus region. However, the neutrino emission from the entire Cygnus Bubble remains unclear.}


In this study, we conduct two analyses on the Cygnus Bubble using a neutrino data sample of IceCube track events from 2011 to 2018. Firstly, we search for neutrino emission from the Cygnus Bubble with a template likelihood method and set 90\% C.L. upper limits on the muon neutrino flux. In addition to using the $\gamma$-ray flux maps as the neutrino emission template, we employ six other templates for comparison, testing different template radii.
Secondly, we scan the region of the Cygnus Bubble and obtain neutrino hotspots, which are compared, respectively, with $\gamma$-ray hotspots, gas distribution, and sources from the TeVCat\footnote{http://tevcat.uchicago.edu} \citep{2008ICRC....3.1341W} and the first LHAASO catalog \citep{LHAASO:2023rpg}.

This paper is organized as the following structure. The LHAASO observations and the IceCube muon-track data are introduced in the \autoref{sec:data}. The analysis methods for the template search and Cygnus Bubble scan are introduced in \autoref{sec:method}. Results and further discussions are shown in \autoref{sec:result}. Finally, \autoref{sec:conclu} summarizes the conclusions. 

\section{Neutrino and Gamma-Ray Data} \label{sec:data}
\subsection{IceCube Neutrino Sample}

The IceCube Neutrino Observatory is a cubic kilometer detector located at the South Pole \citep{IceCube:2008qbc}. Installed between $1.45$ and $2.45~\mathrm{km}$ below the surface of ice, IceCube consists of 86 strings equipped with digital optical modules (DOMs), which can detect Cherenkov light emitted by the secondary charged particles \citep{IceCube:2016zyt}. Muon neutrinos propagating inside the Earth can produce ultrarelativistic muons via charge-current (CC) interactions. These muons, when traversing the detector, will leave a track-like signature.
With high statistics and a typical angular resolution of $\lesssim~1^{\circ}$ at $\sim \mathrm{TeV}$ \citep{IceCube:2019cia}, track-like events are adequately used for neutrino source searches. 

In the analyses, we use 7 years of all-sky muon track data collected by the completed $86$-string detector, namely IC86-2011 (IC86-I) and IC86-2012-18 (IC86-II) \citep{icpstrack2018}. 
The data consists of three components: (i) Experimental data events, including the reconstructed direction with the R.A. ($\alpha$) and decl. ($\delta$), angular uncertainty ($\sigma$), and reconstructed muon energy ($E_{\rm rec}$) for each event. (ii) Instrument response functions, including the effective area $A_{\rm eff}(E_{\nu},\delta_{\nu})$ and the smearing function $M(E_{\rm rec}|E_{\nu},\delta_{\nu})$.
The smearing function gives the fraction count of simulated signal events within the reconstructed muon energy ($E_{\nu}$, $\delta_{\nu}$, $E_{\rm rec}$) bin relative to all events in the ($E_{\nu}$, $\delta_{\nu}$) bin. With the smearing function, the signal energy probability density function (PDF) of the likelihood can be derived under a source spectrum assumption. (iii) The detector uptime, which records the periods of data taking.

\subsection{LHAASO $\gamma$-Ray Data}
LHAASO comprises composite detection arrays that aim to study cosmic rays and $\gamma$-rays \citep{Zhen2010AFP}. Located at $\sim 29^{\circ}$ North in Sichuan Province, China, LHAASO covers a large sky region spanning from $-21^{\circ}$ to $79^{\circ}$ in declination.
Two arrays of LHAASO, the Kilometer Square Array (KM2A) and the Water Cherenkov Detector Array (WCDA), are used for $\gamma$-ray detection. 
The $\sim 1.3~\mathrm{km^2}$ KM2A is able to detect photons with energies from $10~\mathrm{TeV}$ to several $\mathrm{PeV}$, while the $0.078~\mathrm{km}^2$ WCDA probes lower energy photons ranging from $100~\mathrm{GeV}$ to $20~\mathrm{TeV}$ \citep{hehh:2018}. The angular resolution of KM2A is $0.4^{\circ}$ at $30~\mathrm{TeV}$ and can reach $0.2^{\circ}$ at $1~\mathrm{PeV}$
\citep{LHAASO:2019qtb}. For the WCDA, the angular resolution is better than $0.2^{\circ}$ at 10 TeV.

The Cygnus Bubble, recently reported by LHAASO, was measured by both KM2A and WCDA \citep{cygnusbubble2023}. 
The residual structure extends to $\sim 10^{\circ}$ after the removal of all resolved $\gamma$-ray sources and applying a circular mask with a radius of $2.5^{\circ}$ around LHAASO J2018+3651. The $\gamma$-ray excess within a radius of $6^{\circ}$ is still clear after accounting for the diffuse $\gamma$-ray background.
In the $6^{\circ}$ radius region, the energy spectrum of Cygnus Bubble is fitted by a log-parabola function with energies from $2~\mathrm{TeV}$ to $2~\mathrm{PeV}$. The fitted \textcolor{black}{photon index of Cygnus Bubble is $\Gamma = (2.71\pm 0.02)+(0.11\pm 0.02)\times {\rm log}_{10}(E/10~\mathrm{TeV})$. Eight photons with energies above 1 PeV are detected in this region, indicating the existence of super PeVtron(s).
The significance maps of different energy bands show a brightening in the center associated with massive molecular clouds.}
The $\gamma$-rays from the Cygnus Bubble are characterized by four components: two diffuse components with the $\gamma$-ray emission proportional to the column density of atomic \citep[$\HI$,][]{2016A&A...594A.116H} and molecular clouds \citep[MCs,][]{2001ApJ...547..792D}, and two extended sources, LHAASO J2031+4057 and LHAASO J2027+4119. \textcolor{black}{LHAASO J2031+4057 is only observed by WCDA at energy ranges below 20 TeV, while the other three components are observed both by WCDA and KM2A. The $\gamma$-ray flux map measured by LHAASO can be obtained with the spatial and the spectral information of these components.}

\section{Method}\label{sec:method}

\subsection{Template Search}\label{sec:template_search}

An unbinned maximum likelihood method is widely used in neutrino point source searches \citep{BRAUN2008299,BRAUN2010175}. A detailed description of the point source likelihood can be found in \autoref{appendix:likelihood}. Considering the large extension ($\sim 6^{\circ}$) of the Cygnus Bubble, point source likelihood is not suitable for this analysis. Here, we search for neutrino emission associated with the Cygnus Bubble following the ps-template method \citep{IceCube:2017trr}. 
There are two modifications compared to the point source likelihood. Firstly, a spatial template, rather than a two-dimensional (2D) Gaussian function, is used to describe the spatial distribution of signal events. The ps-template method accounts for the extension of the source by mapping the changing detector acceptance and convolving the template with the angular uncertainty of the events. 
Secondly, unlike in the point source likelihood where the background is estimated using scrambled data with negligible point source signal contribution, for a large extended source, the signal events in the data should be subtracted. Therefore, a signal-subtracted likelihood is constructed, and the background is estimated using scrambled data with the signal contamination subtracted \citep{Pinat:2017ldg}.

The event-wise template likelihood \citep{IceCube:2017trr} is defined as
\begin{equation}
L(n_s,\gamma)  = \prod_{i=1}^{N}\Big(\frac{n_s}{N}S_i(\mathbf{x}_i ,\sigma_i,E_i;\gamma)+\widetilde{D}_i({\rm sin}\delta_i,E_i)-\frac{n_s}{N}\widetilde{S}_i({\rm sin}\delta_i,E_i)\Big),
\end{equation}
\textcolor{black}{where $n_s$ is the number of signal events, and $\gamma$ represents the profile of the neutrino spectrum (e.g., power-law and log-parabola). In the template likelihood, the spectral profile is fixed, and we only fit the number of signal events $\hat{n}_s$ by maximizing the likelihood. The neutrino spectrum is calculated using the parameterized energy distribution of secondary particles produced in p-p interactions, under the assumption that all the $\gamma$-rays originate from hadronic processes (details in \autoref{appendix:connection}). The signal PDF $S_i$ is related to the location $\mathbf{x}_i$, angular uncertainty $\sigma_i$, and muon energy proxy $E_i$ of the $i$-th event. The scrambled background PDF $\widetilde{D}_i$ is obtained from the data and therefore contains the scrambled signal component $\widetilde{S}_i$. Each PDF consists of a spatial term and an energy term.}


The construction of the signal spatial PDF in the template likelihood can be described as 
\begin{equation}
S_i^{\rm spat}(\mathbf{x}_i|\sigma_i,\gamma)  = (T_{\rm spat}(\mathbf{x})\times M_{\rm acc}(\mathbf{x},\gamma)*Gaussian_{\rm 2D}(\sigma_i))(\mathbf{x}_i).
\end{equation}
We start with the Cygnus Bubble template $T_{\rm spat}$, which is treated as the neutrino spatial template. 
By convolving it with IceCube acceptance $M_{\rm acc}$, we can obtain the true neutrino direction after accounting for the detector efficiency.
Then this map is smoothed with a 2D Gaussian of width $\sigma_i$ to account for the angular uncertainty of the events. 
Finally, this map is normalized to unity. 
The background PDF $\widetilde{D}_i$ is constructed as the same method in point source likelihood, while the scrambled signal PDF $\widetilde{S}_i$ is constructed following \citep{Pinat:2017wxs}.

We use the $\gamma$-ray emission and the gas column density to weigh the neutrino emission within a $6^{\circ}$ radius from the bubble center. In the $\gamma$-ray flux template, neutrino emission is assumed to follow the $\gamma$-ray flux map at 7 TeV (WCDA) and 50 TeV (KM2A). In the $\HI$ and MC templates, neutrino emission follows the column densities of $\HI$ and MC, respectively, derived from the $\HI$ and CO emission. In the hydrogen (MC+$\HI$) template, neutrino emission follows the total hydrogen column density ($2N_{\rm H_{2}}+N_{\rm HI}$). In the Gaussian templates, neutrino emission follows the 2D Gaussian distribution with $\sigma=0.33^{\circ}$ for LHAASO J2031+4057 and $\sigma=2.28^{\circ}$ for LHAASO J2027+4119. Finally, we employ a uniform template for comparison with the other templates, assuming a uniform spatial distribution of the neutrino flux.
\textcolor{black}{The above templates cover the region of LHAASO 2018+3651.} In addition to the $6^{\circ}$ radius, we explore templates with radii of $0.7^{\circ}$ and $1.2^{\circ}$ based on recent findings \citep{Neronov:2023hzu}, as well as a $10^{\circ}$ radius according to LHAASO's measurements. The center of each template is LHAASO J2032+4102 ($\mathrm{R.A.}= 308.05^{\circ}$, $\mathrm{decl.}=41.05^{\circ}$).

\subsection{Cygnus Bubble Scan} \label{sec:ps_scan}

We scan a $28^{\circ}\times 22^{\circ}$ region, extending to $\sim 10^{\circ}$ to include the entire bubble, to investigate the relation between the neutrino and $\gamma$-ray hotspots and sources, as well as the distribution of MC and $\HI$ gas.
The region is divided into a grid of points, each occupying an area of $0.1^{\circ} \times 0.1^{\circ}$.
Because the $\gamma$-ray significance maps are smoothed with a Gaussian kernel of $\sigma=0.3^{\circ}$, we scan this region with sources having a matching $\sigma_s=0.3^{\circ}$ extension for consistency.
In the likelihood, the signal spatial PDF for extended source is modified from a 2D Gaussian used for the point source likelihood, as shown in Equation (\ref{2d_gaussian_ps}). The modified signal spatial PDF is defined as 
\begin{equation}
S^{\rm spat} (\mathbf{x}_i|\mathbf{x}_s,\sigma_s,\sigma_i)= \frac{1}{2\pi (\sigma_i^2+\sigma_s^2)} e^{-\frac{|\mathbf{x}_s-\mathbf{x}_i|^2}{2(\sigma_i^2+\sigma_s^2)}},
\end{equation}
where $\sigma_s = 0.3^{\circ}$. Other PDFs remain the same as in the point source likelihood.
For each grid point, we maximize the likelihood (see \autoref{appendix:likelihood}) by fitting two parameters: the number of signal events $\hat{n}_s$ and the spectral index $\hat{\gamma}$ assuming a power-law energy spectrum.




\section{Results and Discussion} \label{sec:result}
\subsection{Template Search Results}

\begin{figure}[htbp]
    \centering
    \includegraphics[width=0.6\textwidth]{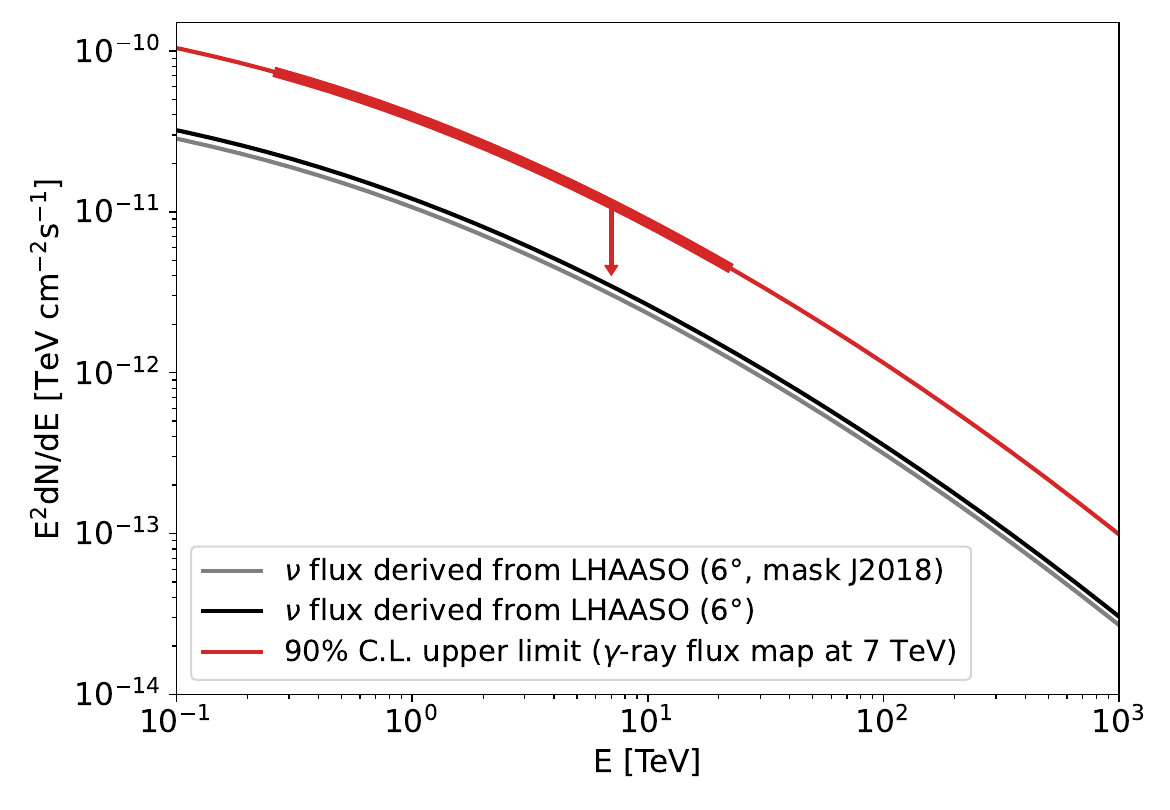}
    \caption{Upper limits ($90\%$ C.L.) on the muon neutrino flux (red) resulting from the template searches using the $\gamma$-ray flux map at 7 TeV for the Cygnus Bubble. \textcolor{black}{The bold red solid line shows the center energy range contributing to $90\%$ significance}. The expected muon neutrino flux (black and gray), derived from LHAASO $\gamma$-ray observations assuming hadronuclear interactions, is also shown. The $2.5^{\circ}$ region centered on LHAASO J2018+3651 is masked for the gray line.} 
    \label{fig_TS_map_upper_limit_E2F} 
\end{figure}

The results of template searches using the $\gamma$-ray flux maps of the Cygnus Bubble are summarized in \autoref{tab_1}. Although some excess from the Cygnus Bubble are observed, the results are not statistically significant. 
\textcolor{black}{The $\gamma$-ray flux template at 7 TeV yields the lower pretrial p-value of 0.176 ($0.9\sigma$). We set upper limits on the muon neutrino flux, which are found to be $\sim$ 3 times higher than the theoretically expected neutrino flux and shown in \autoref{fig_TS_map_upper_limit_E2F}.}
\textcolor{black}{In the previous search in Cygnus region, no neutrino excess ($\hat{n}_s=0$) was found with a pretrial p-value of $0.80$ \citep{2019ICRC...36..932K}, which was probably due to the use of different data samples and templates.}
The results for the other six templates are summarized in Table~\ref{tab_2}. 
\textcolor{black}{Among them, the Gaussian template for LHAASO J2031+4057 ($\sigma=0.33^{\circ}$) yields the lowest pretrial p-value of 0.007 ($2.4\sigma$), while the $\HI$ template yields the largest p-value of 0.291 ($0.6\sigma$).} The lower significant result for the $\gamma$-ray flux map at 50 TeV is probably due to the lack of LHAASO J2031+4057 component.

\begin{figure}[htbp]
    \centering
    \includegraphics[width=0.58\textwidth]{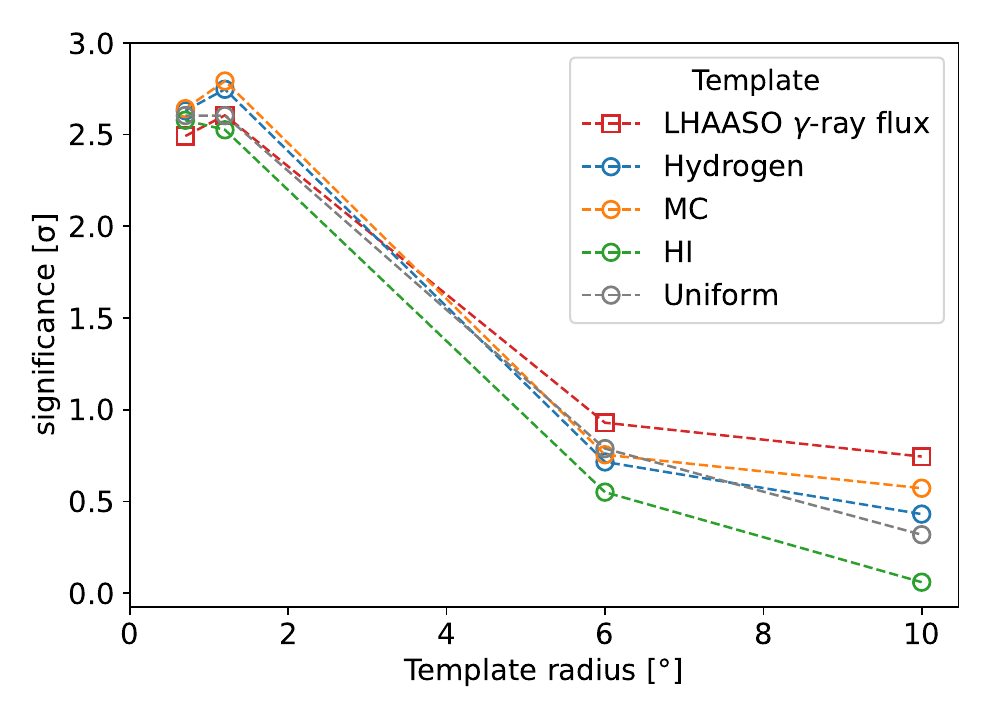}
    \caption{The results (significance) of template searches are shown as a function of the template radius. Results from various templates with radii of $0.7^{\circ}$, $1.2^{\circ}$, $6.0^{\circ}$, and $10.0^{\circ}$ are presented. The results using the LHAASO $\gamma$-ray flux template at 7 TeV are represented as red squares. 
    }
    \label{fig_significance_change_with_radius} 
\end{figure}

The template search results with different template radii of $0.7^{\circ}$, $1.2^{\circ}$, $6.0^{\circ}$, and $10.0^{\circ}$ are shown in Figure~\ref{fig_significance_change_with_radius}. 
The MC template with a radius of $1.2^{\circ}$ gives the most significant result, with a pretrial p-value of $2.6\times 10^{-3}$ ($2.8\sigma$). \textcolor{black}{At larger radii of $6^{\circ}$ and $10^{\circ}$, the neutrino excess of $\gamma$-ray flux template at 7 TeV is more significant. At the radius of $0.7^{\circ}$, the significance of neutrino excess is not sensitive to templates. The best fit number of signal event ($\hat{n}_s=46.6$) for the MC template ($1.2^{\circ}$) is much higher than the expected signal event number ($n_{\rm exp}=4.2$) within the central $1.2^{\circ}$ region.
It seems challenging to attribute all the observed neutrino excess solely to the neutrinos accompanying the $\gamma$-rays from the Cygnus Bubble observed. }

\textcolor{black}{
While the excess of the neutrino signal is not substantial, our results are still consistent with the hadronic origin of the $\gamma$-ray emission from the Cygnus Bubble, as the upper limits of the flux exceed the theoretically predicted neutrino flux.}
To obtain more significant results, additional through-going track events are required. Furthermore, cascade events might be more suitable for measuring neutrinos originating from the extensive region of the Cygnus Bubble. 
This is because the high angular resolution of track events does not provide a significant advantage in reducing background, and cascade events have lower atmospheric background compared to track events.

\begin{table}[htbp]
    \centering
    \begin{tabular}{ccccc}
        \hline
        \hline
        Spatial Template & $\hat{n}_s$ & Upper Limit $\phi_{90\%}$  & Pretrial p-value \\
        \hline
        $\gamma$-ray flux map at $7~\mathrm{TeV}$ & $39.4$ & $5. 69\times 10^{-13}$ &  $0.176$ \\
        $\gamma$-ray flux map at $50~\mathrm{TeV}$  & $29.9$ & $5.31 \times 10^{-13}$ & $0.243$ \\ 
        \hline
    \end{tabular}
    \caption{Results of template searches using the $\gamma$-ray flux templates of the Cygnus Bubble for a $6^{\circ}$ region. The spatial template, the best fit number of signal events $\hat{n}_s$, the $90\%$ C.L. muon neutrino upper limit flux at 5 TeV with units of $\mathrm{TeV}^{-1}\mathrm{cm}^{-2}\mathrm{s}^{-1}$, and the pretrial p-value of each search are listed.}
    \label{tab_1}
\end{table}

\begin{table}[htbp]
    \centering
    \begin{tabular}{ccccc}
        \hline
        \hline
        Spatial Template & $\hat{n}_s$ & Upper Limit $\phi_{90\%}$  & Pretrial p-value \\
        \hline
        MC & $ 31.8 $ & $ 5.32 \times 10^{-13}$  & $0.225 $ \\
        HI  & $ 27.3 $ & $ 5.75 \times 10^{-13}$  & $ 0.291 $ \\
        Hydrogen  & $ 32.2 $ & $ 5.58 \times 10^{-13}$  & $ 0.237 $ \\
        Uniform & $ 41.6 $ & $ 6.68 \times 10^{-13}$  & $ 0.215 $ \\
        LHAASO J2027+4119 ($\sigma=2.28^{\circ}$) & 22.9 & $ 4.61 \times 10^{-13}$  & $ 0.278 $ \\
        LHAASO J2031+4057 ($\sigma=0.33^{\circ}$)  & $ 34.0 $ & $ 3.16 \times 10^{-13}$  & $ 0.007 $ \\
        \hline
    \end{tabular}
    \caption{Results of template searches. Same as Table \ref{tab_1} but for the MC, $\HI$, hydrogen, uniform, and two 2D Gaussian templates within a $6^{\circ}$ radius region. }
    \label{tab_2}
\end{table}

\begin{figure} 
\begin{center}
\includegraphics[width=0.98\linewidth]{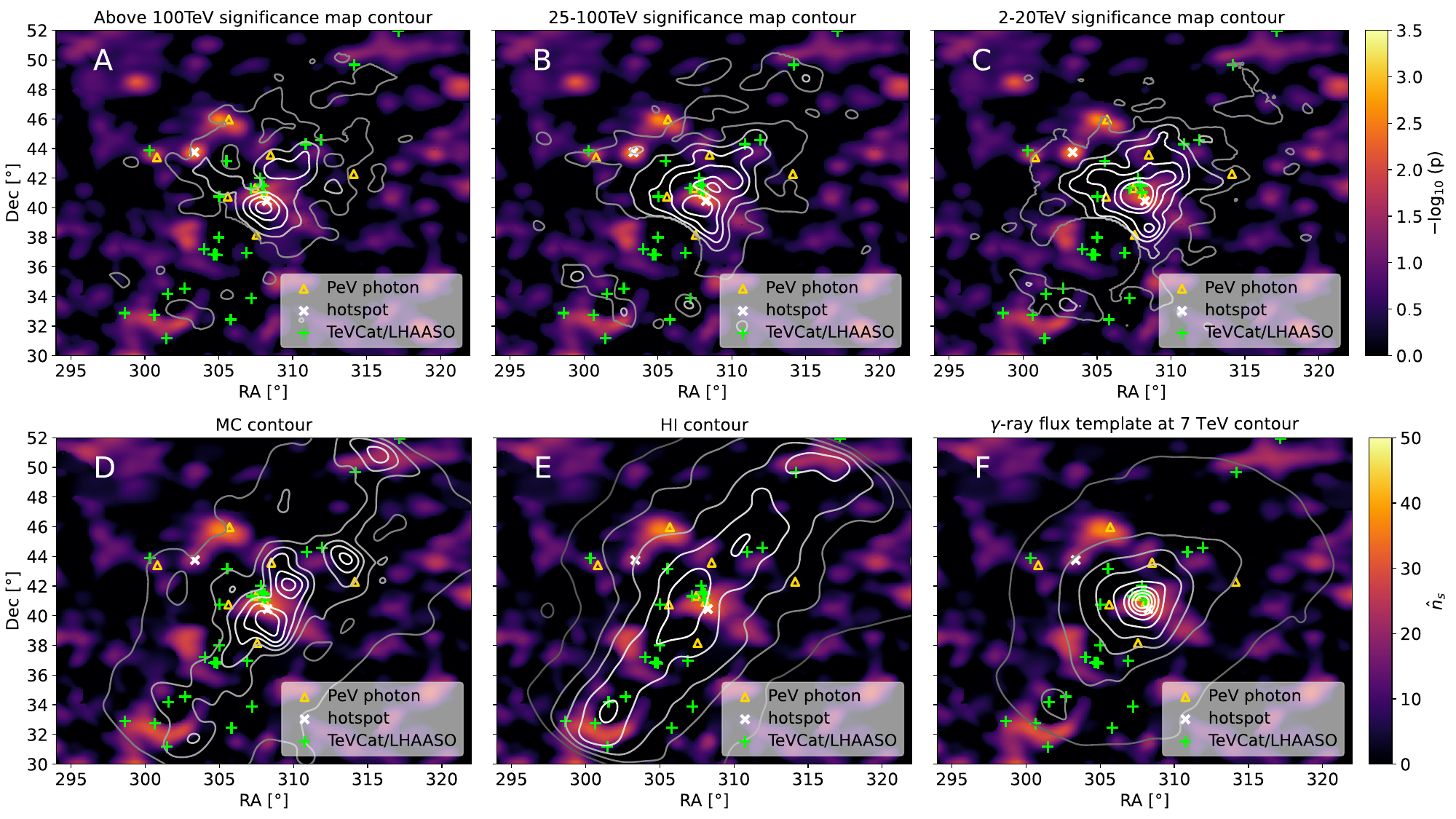}
\end{center}
\caption{Neutrino significance map with the pretrial p-value $-{\rm log_{10}}p$ (panels A-C) and neutrino excess map with the best fit number of signal events $\hat{n}_s$ (panels D-F) for the Cygnus Bubble scan. The $\gamma$-ray significance map with energies above $100~\mathrm{TeV}$ (panel A), $25-100~\mathrm{TeV}$ (panel B), $2-20~\mathrm{TeV}$ (panel C) are shown by contours starting from $3\sigma$ and increasing in steps of $3\sigma$. The spatial distribution of the MC template (panel D), the $\HI$ template (panel E), and the $\gamma$-ray flux template at 7 TeV (panel F) are indicated by contours smoothed with a Gaussian kernel of $\sigma=0.5^{\circ}$ for comparison with the neutrino excess map. Eight photons with energies beyond $1~\mathrm{PeV}$ are shown as gold triangles. Sources from TeVCat and the first LHAASO catalog located in this region are indicated by green plus signs. The most significant point in the entire scan region ($\mathrm{R.A.}=303.35^{\circ}$, $\mathrm{decl.}= 43.75^{\circ}$) and in the central $2^{\circ}$ region ($\mathrm{R.A.}= 308.25^{\circ}$, $\mathrm{decl.} = 40.45^{\circ}$) are indicated by white crosses.}  
\label{significance_map}
\end{figure}



\subsection{Cygnus Bubble Scan Results}


The results of the Cygnus Bubble scan are shown in Figure~\ref{significance_map}, with the upper panels (A-C) illustrating the neutrino significance map and the lower panels (D-F) illustrating the neutrino excess map. In the entire scan region, the most significant point, indicated by the white cross, is found at 
$\mathrm{R.A.} = 303.35^{\circ}$ and $\mathrm{decl.}=43.75^{\circ}$ 
with a pretrial p-value of $2.2 \times 10^{-3}$ $(2.9\sigma)$. This point is located $4.4^{\circ}$ away from the template center, with the best fit parameters being
$\hat{n}_s = 22.2$ and $\hat{\gamma}=2.3$.
In the central $2^{\circ}$ region, the most significant point is found at 
$\mathrm{R.A.} = 308.25^{\circ}$ and $\mathrm{decl.}=40.45^{\circ}$
with a pretrial p-value of 
$6.3 \times 10^{-3}$ $(2.5\sigma)$.
This point is located 
$\sim 0.6^{\circ}$
away from the template center, with the best fit parameters being 
$\hat{n}_s = 31.7 $ and $\hat{\gamma}=4.0$.
\textcolor{black}{We further conduct Monte Carlo simulations by scrambling the $\mathrm{R.A.}$s of IceCube events to compute the post-trial probability of the neutrino hotspot being more significant than the observed one. The post-trial p-value is 
0.96
in the $28^{\circ}\times22^{\circ}$ region and 0.84 (0.18) within the $10^{\circ}$ ($2^{\circ}$) region.} 

The $\gamma$-ray significance map observed by LHAASO partly correlates with the $\HI$ distribution and clearly associates with the dense MC clumps.  If these $\gamma$-rays are produced by cosmic rays interacting with the surrounding gas, the accompanying neutrinos are expected to follow the $\gamma$-ray and gas distributions. The neutrino hotspot in the bubble center is spatially associated with the $\gamma$-ray hotspot below 20 TeV.
However, the neutrino significance map in the larger region (see the upper panels A-C in \autoref{significance_map}) doesn't exhibit an obvious association with the $\gamma$-ray significance map. Similarly, the neutrino excess map (see the lower panels D-F in \autoref{significance_map}) also lacks a clear correlation with the gas distribution or the $\gamma$-ray distribution. 

LHAASO has observed eight PeV photons from the Cygnus Bubble. Two of these PeV photons are associated with the neutrino hotspot in the bubble center. One PeV photon is located close the neutrino excess around 
($\mathrm{R.A.}=305.05^{\circ}$, $\mathrm{decl.}=45.95^{\circ}$), 
and another PeV photon is located close to the neutrino excess around 
($\mathrm{R.A.}=300.55^{\circ}$, $\mathrm{decl.}=43.55^{\circ}$),
which is close to the blazar MAGIC J2001+435 ($\mathrm{R.A.}=300.32^{\circ}$, $\mathrm{decl.}=43.88^{\circ}$). IceCube had previously reported the upper limit on neutrino flux from this source, MG4 J200112+4352, in source-list searches using ten years of track data, yielding a pretrial p-value of 0.21 ($0.8 \sigma$) \citep{IceCube:2019cia}. 

\section{Conclusion} \label{sec:conclu}

In this study, we conducted template searches and a scan of the Cygnus Bubble observed by LHAASO to investigate a potential correlation between neutrinos and $\gamma$-rays using 7 years (2011-2018) of IceCube muon neutrino data observed by the full detector. Using various spatial templates of neutrino emission based on different assumptions, we found no significant neutrino signals in the Cygnus Bubble. 
The most significant result of the template searches is obtained with the MC template in a $1.2^{\circ}$ radius, \textcolor{black}{centered at LHAASO J2032+4102 ($\mathrm{R.A.} = 308.05^{\circ}$, $\mathrm{decl.}=41.05^{\circ}$)}, yielding a pretrial p-value of $2.6\times 10^{-3}\,(2.8\sigma)$. As for the signals from larger regions with radii of $6^{\circ}$ and $10^{\circ}$, the $\gamma$-ray flux template at 7 TeV yields more significant results than other templates.
We obtained $90\%$ C.L. upper limits on the neutrino flux for each template. 
By comparing the resulting upper limits with the theoretically predicted neutrino flux based on $\gamma$-ray observations assuming hadronuclear interactions, we conclude that the neutrino result is consistent with the pure hadronic origin of the $\gamma$-ray emission from the Cygnus Bubble.
\textcolor{black}{The neutrino signals exhibit a stronger tendency to follow the MC distribution compared to the $\gamma$-ray flux distribution in the central region ($\sim 1^{\circ}$) of the Cygnus Bubble.}

\section*{Acknowledgments} \label{sec:ackno}

We thank \textcolor{black}{Andrii Neronov, Teresa Montaruli, and} the LHAASO collaboration for \textcolor{black}{helpful} discussions during the preparation of this work.
D.L.X. and W.L.L. acknowledge the National Natural Science Foundation of China (NSFC) grant (No. 12175137) on “Exploring the Extreme Universe with Neutrinos and multi-messengers” and the Double First Class start-up fund provided by Shanghai Jiao Tong University. T.-Q.H. acknowledges the support of the Special Research Assistant Funding Project of the Chinese Academy of Sciences. H.H.H. acknowledges the support of NSFC grant (No. 12105294) \textcolor{black}{and the Innovative Project of Institute of High Energy Physics (No. E45454U210)}.

\appendix

\section{Likelihood}\label{appendix:likelihood}

A general point source likelihood is defined as
\begin{equation}
L(n_s,\gamma)  = \prod_{i=1}^{N}\Big(\frac{n_s}{N}S_i(\mathbf{x}_i ,\sigma_i,E_i;\mathbf{x}_s,\gamma)+(1-\frac{n_s}{N}){B_i}({\rm sin}\delta_i,E_i)\Big),
\end{equation}
assuming a power law neutrino spectrum $dN_{\nu}/dE_{\nu} \propto E_{\nu}^{-\gamma}$. It should be noted that the neutrino spectrum can be replaced by other functional forms like the log-parabola throughout the likelihood analysis.

In different construction phases of the detector, the effective area and livetime vary. To account for this, we combine each dataset using the likelihood defined as 
\begin{equation}
L = \prod_{k}\prod_{i=1}^{N^k}\Big(\frac{n_s^k}{N^k}S_i+(1-\frac{n_s^k}{N^k}){B_i}\Big),
\end{equation}
where $k$ refers to two data samples, IC86-I and IC86-II.
The number of signal events in the data sample $k$, $n_s^k$, is defined as
\begin{equation}
n_s^k = n_s \times \frac{ t_k \int_0^{\infty} A_{\rm eff}^k(E_{\nu}, \delta_{\nu})E_{\nu}^{-\gamma}dE_{\nu}}{\sum_{k'} t_{k'} \int_0^{\infty}  A_{\rm eff}^{k'}(E_{\nu}, \delta_{\nu})E_{\nu}^{-\gamma}dE_{\nu}},
\end{equation}
where $n_s$ is the total number of signal events, $A_{\rm eff}^k$ and $t_k$ are the effective area and detector livetime of data sample $k$. 

Both signal PDF $S_i$ and background PDF $B_i$ consist of a spatial term and an energy term, which are defined as follows
\begin{equation}
S_i = S^{\rm spat}(\mathbf{x}_i|\sigma_i,\mathbf{x}_s) \times S^{\rm ener}(E_i|\mathbf{x}_s,\gamma),
\end{equation}
\begin{equation}
B_i = B^{\rm spat}(\delta_i) \times B^{\rm ener}(E_i|\delta_i).
\end{equation}
The signal spatial PDF is assumed as a 2D Gaussian, which is constructed as 
\begin{equation}\label{2d_gaussian_ps}
S^{\rm spat} (\mathbf{x}_i|\mathbf{x}_s,\sigma_i)= \frac{1}{2\pi \sigma_i^2} e^{-\frac{|\mathbf{x}_s-\mathbf{x}_i|^2}{2\sigma_i^2}},
\end{equation}
where $\mathbf{x}_i$ and $\mathbf{x}_s$ represent the location of the $i$-th event and the source, and $\sigma_i$ is the reconstructed angular uncertainty of the $i_{th}$ event.
The signal energy PDF is constructed from the smearing function $M(E_{\rm rec}|E_{\nu},\delta_{\nu})$, which gives the reconstructed muon energy proxy at each ($E_{\nu}$, $\delta_{\nu}$) bin. It can be described as
\begin{equation}
S^{\rm ener} (E_i|\mathbf{x}_s,\gamma) = \frac{\sum_j \int_{E_{\nu}^j}^{E_{\nu}^{j+1}}E_{\nu}^{-\gamma}A_{\rm eff}^k(E_{\nu},\delta_s)dE_{\nu}\cdot M(E_i|E_{\nu},\delta_{s})}{\sum_j \int_{E_{\nu}^j}^{E_{\nu}^{j+1}}E_{\nu}^{-\gamma}A_{\rm eff}^k(E_{\nu},\delta_s)dE_{\nu}}
\end{equation}
where j denotes different $E_{\nu}$ bins and $\delta_{s}$ denotes the source declination.

The test statistic (TS) is defined as the ratio of log-likelihood:
\begin{equation}
TS = 2 {\rm ln} \bigg[\frac{L(\hat{n}_s,\hat{\gamma})}{L(n_s=0)} \bigg] =2 \sum_{i}^{N} {\rm ln} \bigg[\frac{n_s}{N}\bigg( \frac{S_i}{B_i}-1 \bigg)+1 \bigg].
\end{equation}



For the signal-subtracted likelihood, the likelihood is modified as
\begin{equation}
L = \prod_{k}\prod_{i=1}^{N^k}\Big(\frac{n_s^k}{N^k}S_i+\widetilde{D}_i-\frac{n_s^k}{N^k}{\widetilde{S}_i}\Big),
\end{equation}
where $\widetilde{S}_i$ denotes the scrambled signal PDF and can be constructed as the projection of signal PDF in declination. 

The TS is modified as
\begin{equation}
TS = 2 {\rm ln} \bigg[\frac{L(\hat{n}_s)}{L(n_s=0)} \bigg] =2 \sum_{i}^{N} {\rm ln} \bigg[\frac{n_s}{N}\bigg( \frac{S_i}{\widetilde{D}_i}-\frac{\widetilde{S}_i}{\widetilde{D}_i} \bigg)+1 \bigg].
\end{equation}
The p-value is the probability of background TS being greater than the observed one.
The distribution of background TS can be derived from the simulations by scrambling the IceCube events in $\mathrm{R.A.}$s. The distribution consists of two parts: the over-fluctuating part with $TS>0$ and the under-fluctuating part with $TS=0$. If the shape of neutrino spectrum is fixed, the over-fluctuating part is expected to follow a $\chi^2_{1}$ distribution, and the fraction of the under-fluctuating part is 0.5.
Thus, the pretrial p-value can be expressed as
\begin{equation}
    p_{\text {value}}=\frac{1}{2}\left[1-\mathop{\mathrm {erf}}\left(\sqrt{{ TS} / 2}\right)\right].
\end{equation}
\textcolor{black}{The neutrino flux upper limit can be derived from the 
$90\%$ C.L. upper limit on the signal event number $n_{90}$, which is defined as 
\begin{equation}
    \frac{\int^{n_{90}}_{0} L(n_s)\pi(n_s)\mathrm{d}n_s}{\int^{N}_{0} L(n_s)\pi(n_s)\mathrm{d}n_s} = 0.9,
\end{equation}\label{eq:ulimit}
where $\pi(n_s)$ is the prior distribution of $n_s$ assumed to be uniform in the range $[0, N]$.}

\section{THE NEUTRINO AND GAMMA-RAY CONNECTION}\label{appendix:connection}

In p-p interactions, the energy spectra of neutrinos and $\gamma$-rays can be expressed as
\begin{equation}
    \phi_{\nu,\gamma}(E) = \int^{\infty}_{E}\frac{{\rm d}E_{p}}{E_{p}}\sigma_{\rm pp}(E_{p})F_{\nu,\gamma}\left(\frac{E}{E_{p}}, E_{p}\right)\int {\rm d}V(\mathbf{x})\frac{cn_{\rm H}}{4\pi d^2}J_{p}(E_{p}, \mathbf{x}),
\end{equation}
where $\sigma_{\rm pp}$ is the cross section of p-p interactions, $F_{\nu}$ ($F_{\gamma}$) is the energy distribution probability of neutrinos ($\gamma$-rays) generated from a cosmic-ray proton with the energy of $E_{p}$ \citep{2006PhRvD..74c4018K}, $V$ is the volume of emission region, $n_{\rm H}$ is the number density of hydrogen, $d$ is the distance to observer, and $J_{p}$ is the energy spectrum of cosmic-ray density. The neutrino spectrum can be obtained from the best fit $\gamma$-ray spectrum $\phi_{\gamma}$ with a photon index of $\Gamma = 2.71 + 0.11 \times {\rm log}_{10}(E_{\gamma}/10~\mathrm{TeV})$, as the integral over the emission region is the same for both neutrino and $\gamma$-ray emissions. The observed neutrinos are assumed to be equal flavor ratio due to neutrino oscillation.

The $\gamma$-ray absorption due to the interstellar radiation field \citep{2017MNRAS.470.2539P} and cosmic microwave background is neglected in the calculation of neutrino spectrum. The attenuation ${\rm exp}(-\tau)$ is $0.99$ for the 100 TeV photon from the direction of Cygnus OB2 with a distance of $1.46\,{\rm kpc}$. Moreover, signal neutrinos from $0.27\,{\rm TeV}$ to $21\,{\rm TeV}$ contribute 90\% of significance for the flux template at 7 TeV (see \autoref{fig_TS_map_upper_limit_E2F}), corresponding to the $\gamma$-rays from $0.54\,{\rm TeV}$ to $42\,{\rm TeV}$ in hadronic origin. Thus, the template search results are not sensitive to the $\gamma$-ray absorption.

\bibliography{sample631}{}
\bibliographystyle{aasjournal}

\end{document}